\documentclass[prl,aps,tightenlines,twocolumn,nofootinbib]{revtex4}
\usepackage{graphicx}

\begin{document}

\title{Sneutrino condensate source for density perturbations, leptogenesis and low reheat temperature}

\author{Anupam Mazumdar~$^{1}$,~and Abdel P\'erez-Lorenzana~$^{2}$}

\affiliation{$^{1}$~CHEP, McGill University, 3600 University Road, Montr\'eal,
Qu\'ebec, H3A 2T8, Canada\\ $^{2}$~Departamento de F\'{\i}sica, Centro
de Investigaci\'on y de Estudios Avanzados del I.P.N.\\
Apdo. Post. 14-740, 07000, M\'exico, D.F., M\'exico\\}

\begin{abstract}
We bring together some known ingredients beyond the Standard Model
physics which can explain the hot Big Bang model with the observed
baryon asymmetry and also the fluctuations in the cosmic microwave
background radiation with a minimal set of assumptions. We propose an
interesting scenario where the inflaton energy density is dumped into
an infinitely large extra dimension. Instead of the inflaton it is the
right handed sneutrino condensate, which is acquiring non-zero vacuum
expectation value during inflation, whose fluctuations are responsible
for the density perturbations seen in the cosmic microwave background
radiation with a spectral index $n_s\approx 1$. The decay of the
condensate is explaining the reheating of the Universe with a
temperature, $T_{rh}\leq 10^{9}$~GeV, and the baryon asymmetry of
order one part in $10^{10}$ with no baryon-isocurvature fluctuations.
\end{abstract}

\maketitle                                                               
Inflation is the most successful paradigm which can explain the scale
invariant density perturbations~\cite{Mukhanov}, the horizon and
flatness problems, however, inflation alone cannot explain the
observable Universe, because inflation leaves the Universe cold and
devoid of (almost) any entropy. It is usually believed that after
inflation the inflaton energy density will be released into the
observable world with the Standard Model (SM) degrees of freedom.
This last but essential point has often been side-lined in many
discussions. The success of inflation lies only if it facilitates the
Big Bang Nucleosynthesis~\cite{Sarkar}.

There are many models of inflation which claim to be satisfying all
the criteria~\cite{Riotto}, inspite of the fact that the inflaton's
identity is largely unknown, and often in the literature it is
regarded as a gauge singlet. Recently with an advent of stringy
motivated inflation, it is possible to realize accelerated expansion
by a geometric stretching of the space time~\cite{Brandenberger1}, or
by inducing potentials from slowly moving D-branes~\cite{Tye}, or via
multiple tachyon condensations~\cite{Panda}.

In all these cases the Universe is assumed to be a four dimensional
surface (brane) embedded in a bulk of $4+n$ dimensions. The $n$ extra
dimensions can be either open~\cite{Sundrum}, or compact~\cite{ADD}
(for a review see Ref.~\cite{rubakov0}). Usually in the brane world
models, the SM fields can be considered to be four dimensional and
localized at the brane, whereas inflation could be driven by either
bulk or brane fields~\cite{Linde}, or due to the modification of the
bulk space time geometry, as it is in the anti-de-Sitter (AdS)
case~\cite{Wands}. All these suggest that there are no dearth of
inflationary models. (for a review on string inspired cosmology,
see~\cite{Quevedo}).

The main aim of this paper is to illustrate a scenario within the
brane world set-up where after inflation the brane is devoid of the
inflaton energy density, therefore the observable world is virtually
cold. In our picture, inflaton plays no role in post-inflationary
cosmology. It is rather the sneutrino condensate which does the whole
job of reheating our Universe, generating primordial density
perturbations and providing baryogenesis via leptogenesis.

Such a scenario is possible if the brane inflaton couples in such a
way that it preferably decays into the bulk degrees of freedom, which
could be either fermions, scalars, or, gravitons. This could happen
for instance, if the inflaton carries some global quantum numbers not
being carried by the brane degrees of freedom, but by some bulk
fields. In this case inflaton energy density can not be dumped into
the SM fields living on the brane.

For simplicity we will work on the 5D set-up, where the bulk is
infinitely large with a warped metric: $ds^2=e^{-k|z|}
g_{\mu\nu}dx^{\mu}dx^{\nu}-dz^2$, with all the SM fields attached to
the brane~\cite{Sundrum}. Here $k$ is a constant that relates the
fundamental five dimensional scale of gravity, $M_5$, with Planck
scale $M_P\sim 2.4\times10^{18}$, by $kM_P^2 = M_5^3$. We will assume
that $k$ is close to the Planck scale to facilitate the standard Hubble
expansion on the brane~\cite{langlois}.

In such a background the wave functions of any bulk fields have a
continuum spectrum~\cite{rubakov0,rubakov}, that starts at $m=0$,
meaning that its Kaluza-Klein (KK) modes can take any momentum along
the extra space coordinate. Now let us assume that the inflaton, which
is a brane scalar field, couples to the bulk degrees of freedom.  In
this case the inflaton can decay into all the continuum KK modes below
its mass, therefore draining its energy from the brane and into the
bulk. Note that the inflaton energy density will be gradually
redshifted into the bulk before becoming vanishingly small at the
brane, this is due to the fact that the 5th dimension is warped such
that the released energy density into the bulk appears to be red
shifted from the brane observer~\cite{Sundrum,rubakov0,rubakov}.

Our scenario can be thought of as a hot radiating plate cooling down
by emitting its energy into a cold surrounding.  It is not hard to see
that this cooling process is extremely efficient.  Let us consider for
instance the coupling, $\phi \bar\psi \psi\delta(y)/M_5$, where $\phi$
is the inflaton of mass $m_\phi$, and $\psi$ is a bulk fermion in an
AdS background. The typical KK expansion of the wave function of the
bulk field goes as $\psi(x,z) = \int (dm/k) \psi_m(x) h_m(z)$, where
at the brane position $h_m(0)\sim\sqrt{m}$, with $m$ being the KK
mass~\cite{rubakov}. The total decay rate of the inflaton into KK
modes is given by
\begin{eqnarray}
\Gamma&=&\int_{0}^{m_\phi}\,dm\,dm'({m_\phi/k^2M_5^2})|h_m(0)h_m'(0)|^2\,,
\nonumber \\
&\approx& \left({m_\phi/M_5}\right)^4 \left({M_P/M_5}\right)^4~ m_\phi\,.
\end{eqnarray}  
Often the inflaton is typically heavy, say about the GUT scale.
Let us assume $M_5\sim 0.1 M_{p}$, then we find the inflaton decay 
rate is quite fast compared to the Hubble expansion rate after the 
end of inflation, e.g. $H\sim m_{\phi}$.

Note, however, that these bulk fields may not be completely harmless.
As they move towards the fifth dimension, their energy density might
eventually collapse to form a black hole at the AdS horizon
\cite{march-russell}. The presence of a Black Hole in the bulk, 
induces a contribution to the brane expansion that looks like a dark
energy contribution which goes as $(r_h/ a(t))^4$, where $a$ is the
brane scale factor and $r_h$ the Black Hole horizon size. The standard
Hubble expansion law, $H^2\sim \rho/M_p^2$, prevails once the energy
density on the brane is below the fundamental scale, which we assume
to hold true always in our case including the inflaton energy
density. In this case, $r_h \sim V(\phi)^{-1/4}$, is much smaller than
the scale factor after the inflaton decay, and therefore such a
potential dark energy contribution, $\rho \sim (1/a^4)$, gives
negligible contribution today.

Once the inflaton energy is redshifted into the bulk (from the brane
point of view), then the challenge is to reproduce the standard hot
Big Bang model, using only brane physics.  In this regard our approach
is simple: a real bang from the sneutrino-cosmology.

The recent advancement made in neutrino experiments point towards the
fact that the neutrinos have non-vanishing masses.  The solar neutrino
deficit is better understood if the electron neutrinos oscillate into
the muon neutrinos controlled by the squared mass difference $\Delta
m^2_{solar}\sim 7\times 10^{-5}~{\rm eV}^2$, with a large mixing
angle, $\tan^2\theta_{solar}\sim 0.5$~\cite{Fukudas}, whereas the
atmospheric neutrino experiments indicate $\nu_{\mu}-\nu_{\tau}$
oscillations with $\Delta m^2_{atm}\sim 2.5\times 10^{-3}~{\rm eV}^2$
and $\sin^2(2\theta)\simeq 1$~\cite{Fukudaa}.

The above required masses are much smaller than those expected if the
neutrinos were Dirac particles, since this would require a fine tuning
of the Dirac Yukawa couplings to one part in $10^{11}$, at least. The
most natural explanation of such small masses, however, comes out if
the neutrinos are Majorana particles, then the small masses can be
understood through the see saw mechanism that involves large right
handed neutrino masses~\cite{Seesaw}.  One advantage of this mechanism
is that the right handed neutrino mass breaks $L$ (or $B-L$) quantum
number, which can be the origin of the observed baryon asymmetry. The
conversion of the lepton asymmetry into the baryons via active SM
sphalerons within a range of $10^{12}~{\rm GeV}\geq T\geq 100$~GeV can
help us producing the observed baryon asymmetry~\cite{Fukugita}. In a
supersymmetric theory, a supersymmetric partner of the right handed
neutrino, the sneutrino, induces leptogenesis, which is in
principle can be tested by the cosmic microwave background radiation
through the baryon-isocurvature fluctuations, e.g~\cite{Mazumdar1}.

The next question is how to reheat and generate the adiabatic density
perturbations? The right handed neutrinos naturally couple to the SM
Higgs and the lepton doublets, therefore they decay into the SM
degrees of freedom. Indeed, as we promote our idea to supersymmetry,
the sneutrino can decay into the Higgsinos and the sleptons. The
lightest stable supersymmetric particle can easily account for the
observed cold dark matter~\cite{Olive}. During inflation the sneutrino
condensate obtains a non-vanishing vev, eventhough its energy density
during inflation is subdominant, but after inflation, once the
inflaton energy has been released away into the bulk, the sneutrino
condensate decays and becomes the only source for the entropy
production.  Therefore once inflation is over, the lightest sneutrino
condensate, whose mass is lighter than the Hubble expansion rate
during inflation, starts oscillating coherently and eventually
decaying into the SM degrees of freedom~\cite{Marieke}.

Next comes the challenge for generating the adiabatic density
perturbations, which happens naturally when the sneutrino condensate
decays. During inflation the sneutrino condensate generates quantum
fluctuations which are stretched outside the horizon. These
perturbations are known as the isocurvature fluctuations, however,
once inflation ends the dominant energy density is in the sneutrino
condensate, which when decays, by converting its entire isocurvature
perturbations to the adiabatic ones.  This process is known as a
curvaton scenario~\cite{Sloth}, its supersymmetric implementation has
been given in Ref.~\cite{Enqvist}.

For the purpose of illustration, let us assume a simple superpotential
for the sneutrino condensate
\begin{equation}
W=({1}/{2})M_{N} {\bf N}~{\bf N}+h{\bf N}~{\bf L}~{\bf H_{u}}\,,
\end{equation}
where ${\bf N},~{\bf L},~{\bf H_{u}}$ stand for the neutrino, the
lepton and the Higgs doublet. We have assigned an odd $R$-parity for
the right handed sneutrinos. The only possible interactions with
leptons and the Higgs fields are through the Yukawa matrix,
$h_{i\ell}$, where $~i=1,2,3$ and $\ell=e,\mu,\tau$ respectively. The
superpotential term induces a potential, $V=M^2_{N}{\widetilde N}^2$,
where $\widetilde N$ denotes the sneutrino. For simplicity we always
assume a diagonal basis for the right handed neutrino mass matrix. The
left handed neutrinos obtain masses via the see-saw mechanism,
$m_{\nu}\sim m_{D}^\dagger M_{N}^{-1}m_{D}$, where $m_{D}$ is the Dirac
mass matrix.

The lightest right handed sneutrino, whose mass is smaller than the 
Hubble expansion rate during inflation, can act as a cosmic condensate
\cite{Mazumdar1}. The heavier sneutrinos would roll down to the bottom of the
potential during inflation. They will also generate isocurvature
fluctuations, but their amplitude will be suppressed compared to the
lighter ones~\cite{LM,Mazumdar1}. In this letter we will ignore their
dynamics. We will only deal with the lightest sneutrino which we
identify as the right handed electron sneutrino ($N_1$) that we will
just call hereafter $N$, and $M_N$ to its mass. We also assume that
its largest Yukawa coupling is that of $\tau$ doublet, that we take
$h_{1\tau}\sim 10^{-4}-10^{-5}$. Note that by taking $M_{N}\sim
10^{10}$~GeV, with $M_{N2,3}\gg M_{N}$, we could obtain a small
electron-tau neutrino mixing in the left handed sector, as it seems to
be required by small $\theta_{13}$ angle in the neutrino experiments.

The important point is that the isocurvature perturbations seed the
adiabatic ones on the largest scales~\cite{Mukhanov}.
Nevertheless, the amplitude of such fluctuations must be of
the correct order of magnitude during inflation, which then requires
\begin{equation}
    V''({\widetilde N}_*) = M^2_{N}\sim \alpha^2 H_*^2, \qquad
    {H_*}/{{\widetilde N}_*}=\delta\,,
\end{equation}
where star denotes the value evaluated at the horizon crossing,
$\alpha \ll 1$, and the perturbation, $\delta \sim 10^{-5}$. On the
other hand the scale of inflation can be parameterized by
$V_{I}^{1/4}\sim(H_*M_p)^{1/2}$. The fluctuation in the sneutrino
condensate can be imprinted upon radiation after its decay.  In this
terms, the spectral index of microwave temperature perturbations can
then be evaluated as
\begin{equation}
\label{nspectr}
    n_s-1 = 2 {\dot{H_*}}/{H_*^2}
    + ({2}/{3})({M_{N}}/{H_*^2})\sim ({2}/{3})\alpha^2\,.
\end{equation}
For  $\alpha \sim 10^{-1}$, we find that $n_s$ is be fairly close
to one for $\dot H_{\ast}/H^2_{\ast}\approx 0$, which is consistent 
with the recent WMAP observation of $n_s=0.99 \pm 0.04$ \cite{Spergel}.

The perturbative decay rate of the lightest sneutrino condensate can
be estimated by $\Gamma\sim |h_{1\tau}|^2M_{N}/4\pi$, and the reheat
temperature, $T_{rh}\sim 0.1\sqrt{\Gamma M_{p}}$. We obtain
\begin{equation}
T_{rh}\sim 10^{9}\left({h_{1\tau}}/{10^{-4}}\right)\left({M_{N}}/{
10^{10}}\right)^{1/2}~{\rm GeV}\,.
\end{equation}
Note that the reheat temperature is low enough to avoid the gravitino
problem (reheat temperature below $10^{9}$~GeV) \cite{Ellis}. The
bound on the lightest right handed neutrino mass has to be around
$10^{10}$~GeV, which sets the scale of inflation to be $H_{\ast}\sim
10^{10}/\alpha$~GeV. If $\alpha\sim 10^{-1}$, then $H_{\ast}\sim
10^{11}$~GeV and ${\widetilde N}\sim 10^{16}$~GeV, close to the grand
unification scale. Note that we ignored the D-term contributions in
the sneutrino potential, however, it is safe to assume so if the
sneutrino vev is smaller than the scale at which the gauge group which
embodies right handed neutrino breaks down, e.g. $SO(10)$ gauge
group. Note also, that the right handed neutrino mass scale is
different from the GUT scale, since it can actually come via some non
renormalizable operators, as it happens in many $SO(10)$
constructions.

Let us discuss baryogenesis. The sneutrino decay also induces lepton
asymmetry due to the CP violation. The CP asymmetry in our scenario
can be calculated by computing the interference between the tree-level
and the one loop diagrams of $\widetilde N$ going into 
$l_{\ell}{\widetilde H_{u}}$ and the antislepton 
$l_{\ell}^{\ast}H_{u}^{\ast}$. The CP asymmetry is given
by~\cite{Fukugita,Murayama}
\begin{equation}
\epsilon\sim ({\ln 2}/{8\pi}){{\cal I}m h_{3\tau}^{\ast~2}}\,,
\end{equation}
where we assumed  $h_{1\tau}\geq h_{1e},~h_{1\mu}$, with  $h_{3\tau}$
dominance, and both $h_{1\tau}$ and $M_N$ are real.

Now we can predict the overall light neutrino mass scale with a help
of seesaw formula, we get
$m_{\nu,\tau}=|h_{3\tau}|^2{\sin^2\beta}/{(2\sqrt{2}G_{F}M_{N,3})}$,
where $G_{F}$ is the Fermi constant, $M_{N,3}$ is the mass of the
right handed tau sneutrino, and $\tan\beta=\langle
H_{u}\rangle/\langle H_{d}\rangle$.  If we take $|h_{3\tau}|\sim
10^{-2}$, with a typical $\tan\beta \sim 10$ and $M_{N,3}\sim
10^{11}$~GeV we obtain $m_{\nu,\tau}\sim 0.3$~eV. Further note that
this value is already at the desired scale for the right handed
neutrino masses. Also a slightly smaller Yukawa coupling or a larger
$M_{N3}$ can easily bring the mass scale down without affecting other
predictions.

The lepton asymmetry is converted by the SM sphalerons, as 
a result the net baryon asymmetry is given by~\cite{Murayama}
\begin{equation}
{n_{B}}/{s} \sim \epsilon ({8}/{15})({T_{rh}}/{M_{N}})\,.
\end{equation}
The ratio $T_{rh}/M_{N}$ arises due to the entropy dilution.  We note
that with $\epsilon\sim 10^{-8}$, we can easily generate the baryon
asymmetry of order one part in $10^{10}$ for $T_{rh}\sim 10^{9}$~GeV
and $M_{N}\sim 10^{10}$~GeV. The actual prediction is an interplay
between $T_{rh}$, $M_{N}$ and $\epsilon$, but it is interesting to
see that it is possible to obtain the right number without much fine tuning.

Also note that the fluctuations in the reheat temperature leads to the
fluctuations in the baryon asymmetry~\cite{Mazumdar1},
$\delta(n_{B}/s)\propto \delta T_{rh}$, however, in our case it is
possible to show that there is no baryon isocurvature fluctuations
generated due to the sneutrino decay. The fact that the sneutrino
completely dominates the energy density while decaying, therefore
converting all its isocurvature fluctuations into the adiabatic
ones. In our simple set-up we strictly predict $S_{B}=0$.

To summarize, in our set-up, we have completely relaxed the inflaton
sector, although inflation certainly solves the horizon and the
flatness problems, however, the inflaton energy density need not
reheat the Universe with the SM degrees of freedom. We showed that in
a brane world set up it is possible that the inflaton energy density
can be red shifted away into the bulk. The density perturbations,
reheating the Universe with SM degrees of freedom, and the baryon
asymmetry are all served by a single source; the right handed Majorana
neutrino sector within supersymmetry. The neutrino sector can be
embedded in a grand unified gauge group such as in
$SO(10)$. Interesting points to note that in our case the sneutrino
vev is consistent with the $SO(10)$ breaking scale, the reheat
temperature is low enough to avoid the gravitino problem, the spectral
index is close to one, and there is no baryon-isocurvature
fluctuations.

A.M. is a CITA-National fellow and acknowledges Kari Enqvist, Jim
Cline, Guy Moore and Horace Stoica for helpful discussion.

\vskip10pt



\begin{thebibliography}{90}


\bibitem{Mukhanov}
V. F. Mukhanov, H. A. Feldman, and R. H. Brandenberger, 
Phys. Rept. {\bf 215}, 
203 (1992).

\bibitem{Sarkar}
S. Sarkar, Rep. Prog. Phys. {\bf 59}, 1493 (1996);
K. A. Olive, G. Steigman, and T. P. Walker, Phys. Rep. {\bf 333}, 389 (2000).

\bibitem{Riotto}
D. H. Lyth, and A. Riotto, Phys. Rept. {\bf 314}, 1 (1998).

\bibitem{Brandenberger1}
R.~Brandenberger, D.~A.~Easson and A.~Mazumdar,
arXiv:hep-th/0307043.

\bibitem{Tye}
G.~R.~Dvali and S.~H.~Tye, Phys.\ Lett.\ B {\bf 450}, 72 (1999);
C.~P.~Burgess, M.~Majumdar, D.~Nolte, F.~Quevedo, G.~Rajesh and R.~J.~Zhang,
JHEP {\bf 0107}, 047 (2001);J.~Garcia-Bellido, R.~Rabadan and F.~Zamora,
JHEP {\bf 0201}, 036 (2002);
G.~Shiu and S.~H.~Tye,
Phys.\ Lett.\ B {\bf 516}, 421 (2001);
N.~Jones, H.~Stoica and S.~H.~Tye,
JHEP {\bf 0207}, 051 (2002);
K.~Dasgupta, C.~Herdeiro, S.~Hirano and R.~Kallosh,
Phys.\ Rev.\ D {\bf 65}, 126002 (2002);B.~s.~Kyae and Q.~Shafi,
Phys.\ Lett.\ B {\bf 526}, 379 (2002);
S.~Kachru, R.~Kallosh, A.~Linde, J.~Maldacena, L.~McAllister and S.~P.~Trivedi,
arXiv:hep-th/0308055.

\bibitem{Panda}
A.~Mazumdar, S.~Panda and A.~Perez-Lorenzana,
Nucl.\ Phys.\ B {\bf 614}, 101 (2001);


\bibitem{Sundrum}
L. Randall and R. Sundrum, Phys. Rev. Lett.~{\bf 83}, 4690 (1999).

\bibitem{ADD}
N.Arkani-Hamed, S. Dimopoulos, and G. Dvali, Phys. Lett {\bf B 429},
263 (1998);\prd {\bf 59}, 086004 (1999); I. Antoniadis, N.
Arkani-Hamed, S. Dimopoulos, and G. Dvali, Phys. Lett. {\bf B 436},
257 (1998).

\bibitem{rubakov0}
V.~A.~Rubakov,
Phys.\ Usp.\  {\bf 44}, 871 (2001)
[Usp.\ Fiz.\ Nauk {\bf 171}, 913 (2001)]

\bibitem{Linde}
N. Kaloper and A. Linde, Phys. Rev. D {\bf 59}, 101303 (1999);
 A. Mazumdar, Phys. Lett. B {\bf  469}, 55 (1999);
 N. Arkani-Hamed, {\it et al.}, Nucl. Phys. B{\bf 567}, 189 (2000).
 R.N. Mohapatra, A. P\'erez-Lorenzana and C.A. de S. Pires,
 Phys. Rev. {\bf D 62}, 105030 (1999);
 A.~Mazumdar and A.~Perez-Lorenzana, Phys.\ Lett.\ B {\bf 508}, 340 (2001);
 A.~M.~Green and A.~Mazumdar, Phys.\ Rev.\ D {\bf 65}, 105022 (2002);
 A.~Mazumdar, R.~N.~Mohapatra and A.~P\'erez-Lorenzana, arXiv:hep-ph/0310258.


\bibitem{Wands}
R.~Maartens, D.~Wands, B.~A.~Bassett and I.~Heard, Phys.\ Rev.\ D {\bf
62}, 041301 (2000);A.~Mazumdar and J.~Wang, Phys.\ Lett.\ B {\bf 490},
251 (2000);E.~E.~Flanagan, S.~H.~Tye and I.~Wasserman, Phys.\ Rev.\ D
{\bf 62}, 044039 (2000); A.~Mazumdar, Phys.\ Rev.\ D {\bf 64}, 027304
(2001); A.~Mazumdar, Nucl.\ Phys.\ B {\bf 597}, 561 (2001).


\bibitem{Quevedo}
F.~Quevedo, Class.\ Quant.\ Grav.\  {\bf 19}, 5721 (2002).

\bibitem{langlois}
P. Binetruy, C. Deffayet, D. Langlois, Nucl. Phys. {\bf B565}, 269 (2000); 
J.M. Cline, C. Grojean, G. Servant, \prl {\bf 83}, 4245 (1999). 


\bibitem{rubakov}
S.L. Dubovsky, V.A. Rubakov and P.G. Tinyakov, \prd {\bf 62}, 105011 (2000).


\bibitem{march-russell}
A. Hebecker and J. March-Russell,  Nucl. Phys. B {\bf 608}, 375 (2001). 

\bibitem{Fukudas}
S. Fukuda et. al. [Super-Kamiokanade Collaboration], Phys. Rev. Lett.
{\bf 86}, 5656 (2001); Q. R. Ahmad et.al. [SNO Collaboration],
Phys. Rev. Lett. {\bf 87}, (2001).

\bibitem{Fukudaa}
S. Fukuda et. al. [Super-Kamiokanade Collaboration], Phys. Rev. Lett. {\bf 85},
3999 (2000).

\bibitem{Seesaw}
M. Gell-Mann, P. Ramond, and R. Slansky, in {\it Supergravity},
eds. P. van Niewenhuizen and D.Z. Freedman (North Holland 1979);
T. Yanagida, Proceedings of {\it Workshop on Unified Theory and
Baryon number in the Universe}, eds. O. Sawada and A. Sugamoto (KEK 1979);
R.N. Mohapatra, and G. Senjanovi{\'c}, Phys. Rev. Lett. {\bf 44}, 912
(1980).

\bibitem{Fukugita}
M. Fukugita, and T. Yanagida, Phys. Lett. B {\bf 174}, 45 (1986);
M.~Fukugita and T.~Yanagida, Phys.\ Rev.\ D {\bf 42}, 1285 (1990).

\bibitem{Mazumdar1}
A.~Mazumdar, arXiv:hep-ph/0306026;
A.~Mazumdar, Phys.\ Lett.\ B {\bf 580}, 7 (2004).

\bibitem{Olive}
J.~R.~Ellis, {\it et. al.}, Phys.\ Lett.\ B {\bf 565}, 176 (2003).


\bibitem{Marieke}
M.~Postma and A.~Mazumdar, JCAP {\bf 0401} 005, (2004).


\bibitem{Sloth}
K.~Enqvist and M.~S.~Sloth, Nucl.\ Phys.\ B {\bf 626}, 395 (2002)
D.H. Lyth and D. Wands, Phys. Lett. B {\bf 524}, 5 (2002); T.~Moroi
and T.~Takahashi, Phys.\ Lett.\ B {\bf 522}, 215 (2001)
[Erratum-ibid.\ B {\bf 539}, 303 (2002)].


\bibitem{Enqvist}
K.~Enqvist, S.~Kasuya and A.~Mazumdar, Phys.\ Rev.\ Lett.\ {\bf 90},
091302 (2003); K.~Enqvist, A.~Jokinen, S.~Kasuya and A.~Mazumdar,
Phys. Rev. D {\bf 68}:103507, (2003). M. Postma, Phys.\ Rev.\ D {\bf
67}, 063518 (2003); K.~Hamaguchi, {\it et. al.}, arXiv:hep-ph/0308174;
K.~Hamaguchi, {\it et. al.}, Phys.\ Rev.\ D {\bf 65}, 043512
(2002);J.~McDonald, Phys.\ Rev.\ D {\bf 68}, 043505 (2003);
K.~Enqvist, S.~Kasuya and A.~Mazumdar, arXiv:hep-ph/0311224;
K. Enqvist, A. Mazumdar and A. Perez-Lorenzana, arXiv:hep-th/0403044.


\bibitem{LM}
A. D. Linde, and V. F. Mukhanov, Phys. Rev. D {\bf 56}, 535 (1997);
A. R. Liddle, and A. Mazumdar, Phy. Rev. D {\bf 61}, 123507 (2000).


\bibitem{Spergel}
  Spergel {\it et al.}, astro-ph/0302209.


\bibitem{Ellis}
J. Ellis, {\it et al.}, Phys. Lett. B {\bf 145},
181 (1984); A. L. Maroto and A. Mazumdar, Phys. Rev. Lett. {\bf 84},
1655 (2000); R. Kallosh {\it et al.}, Phys. Rev. D {\bf 61}, 103503
(2000); R.~Allahverdi, {\it et al.}, Phys.\ Rev.\ D {\bf 64}, 023516 (2001).

\bibitem{Murayama}
H. Murayama, H. Suzuki, T. Yanagida and J. Yokoyama, Phys. Rev. Lett.
{\bf 70}, 1912 (1993).






\end{thebibliography}
\end{document}